\documentclass[apj]{emulateapj}

\usepackage{multirow}
\usepackage{amsmath}
\usepackage{color}

\usepackage{hyperref}	
\hypersetup{colorlinks=true,linkcolor=blue,citecolor=blue,filecolor=blue,urlcolor=blue}

\begin{document}


\title{The Impact of Modeling Assumptions in Galactic Chemical Evolution Models}

\author{Benoit C\^ot\'e,\altaffilmark{1,2,5,6} Brian W. O'Shea,\altaffilmark{2,3,4,5} Christian Ritter,\altaffilmark{1,5,6} 
Falk Herwig,\altaffilmark{1,5,6} Kim A. Venn\altaffilmark{1}}

\altaffiltext{1}{Department of Physics and Astronomy, University of Victoria, Victoria, BC, V8W 2Y2, Canada}
\altaffiltext{2}{National Superconducting Cyclotron Laboratory, Michigan State University, East Lansing, MI, 48824, USA}
\altaffiltext{3}{Department of Physics and Astronomy, Michigan State University, East Lansing, MI, 48824, USA}
\altaffiltext{4}{Department of Computational Mathematics, Science and Engineering, Michigan State University, East Lansing, MI, 48824, USA}
\altaffiltext{5}{Joint Institute for Nuclear Astrophysics - Center for the Evolution of the Elements, USA}
\altaffiltext{6}{NuGrid Collaboration, \href{http://www.nugridstars.org}{http://www.nugridstars.org}}

\label{firstpage}

\begin{abstract}

We use the OMEGA galactic chemical evolution code to investigate how the assumptions used for the treatment of galactic inflows and outflows impact numerical predictions.  The goal is to determine how our capacity to reproduce the chemical evolution trends of a galaxy is affected by the choice of implementation used to include those physical processes.  In pursuit of this goal, we experiment with three different prescriptions for galactic inflows and outflows and use OMEGA within a Markov Chain Monte Carlo code to recover the set of input parameters that best reproduces the chemical evolution of nine elements in the dwarf spheroidal galaxy Sculptor.  This provides a consistent framework for comparing the best-fit solutions generated by our different models.  Despite their different degrees of intended physical realism, we found that all three prescriptions can reproduce in an almost identical way the stellar abundance trends observed in Sculptor.  This result supports the similar conclusions originally claimed by \cite{rs13} for Sculptor.  While the three models have the same capacity to fit the data, the best values recovered for the parameters controlling the number of Type~Ia supernovae and the strength of galactic outflows, are substantially different and in fact mutually exclusive from one model to another.  For the purpose of understanding how a galaxy evolves, we conclude that only reproducing the evolution of a limited number of elements is insufficient and can lead to misleading conclusions.  More elements or additional constraints such as the galaxy's star formation efficiency and the gas fraction are needed in order to break the degeneracy between the different modeling assumptions.  Our results show that the successes and failures of chemical evolution models are predominantly driven by the input stellar yields, rather than by the complexity of the galaxy model itself.  Simple models such as OMEGA are therefore sufficient to test and validate stellar yields.  OMEGA is part of the NuGrid chemical evolution package and is publicly available online at \href{http://nugrid.github.io/NuPyCEE}{http://nugrid.github.io/NuPyCEE}.
\\
\end{abstract}

\begin{keywords}
{galaxies: abundances -- galaxies: evolution -- galaxies: dwarf -- stars: yields}
\end{keywords}

\section{Introduction}
\label{sect_intro}
Large spectroscopic stellar surveys, such as SEGUE (e.g., \citealt{y09}), RAVE (e.g., \citealt{k13}), and APOGEE (e.g., \citealt{a14}), offer an opportunity to improve our understanding of the formation and evolution of the elements in the Milky Way and its satellite galaxies.  These modern surveys combine several key attributes -- a large number of stars, a coherent selection that covers different galactic environments, and a uniform data reduction and analysis.  Coupled with wide-sky-area photometric surveys and the kinematics of local stellar populations, modern spectroscopic surveys will provide valuable information about the origin and the evolution of the Milky Way.  The breadth and precision of modern stellar observations require the use of statistical techniques and sophisticated theoretical models to untangle the complex relationship between chemistry and galaxy evolution.  A robust understanding of the uncertainties inherent in those models is necessary in order to provide a measure of confidence in the resulting predictions, and to understand which questions can be reliably answered by the available data and models.

This paper is part of a series that presents a numerical framework for modeling the chemical evolution of galaxies and for evaluating the reliability of its numerical predictions (see \citealt{c_nic_16}).  Our motivation is to bridge the fields of nuclear physics, stellar evolution, galaxy evolution, cosmological structure formation, and modern statistical methods.  This framework consists of a chain of models that so far includes the NuGrid stellar evolution models which use the JINA ReacLib nuclear reaction network (\citealt{p13}; Ritter et al.~in prep.), the SYGMA  (Stellar Yields for Galactic Modeling Applications, Ritter et al.~in prep.) simple stellar population model, the OMEGA (One-zone Model for the Evolution of GAlaxies, the present paper) single-zone galactic chemical evolution model, and the STELLAB\footnote{See \href{http://nugrid.github.io/NuPyCEE}{http://nugrid.github.io/NuPyCEE} for an introduction.} (STELLar ABundances) module to plot observational data.  We are currently extending our framework to model the hierarchical growth of galaxies with merger trees taken from cosmological simulations.  One of our goals is to establish the predictive power of our tools by quantifying how the uncertainties inherent in the different components of our framework accumulate and propagate in our chemical evolution predictions.  We also plan to use this framework to test nuclear astrophysics and stellar modeling assumptions.

In \cite{c15}, we considered a one-zone closed-box model and used a Monte Carlo approach to quantify the resulting scatter caused by the uncertainties in fundamental input parameters, which were mostly associated with the modeling of simple stellar populations (SSPs).  Among the seven input parameters explored in that study, we found that the slope of the high-mass end of the stellar initial mass function and the number of SNe~Ia per M$_\odot$ formed are currently generating the most uncertainties in our predictions.  We refer to \cite{g02}, \cite{rcmt05,rktm10}, \cite{msrv09}, \cite{w09}, \cite{y13}, and \cite{mcgg15} for complementary studies regarding uncertainties in galactic chemical evolution models.  In this paper, we focus on the impact of modeling assumptions in predicting the chemical evolution of local dwarf spheroidal galaxies, using one-zone open-box models.

Galactic inflows and outflows play a significant role in shaping the chemical evolution of galaxies (e.g., \citealt{t80,p08,m14}).  Introducing inflows dilutes the metal content of galactic gas and allows one to start a model with a smaller gas reservoir,  increasing the speed of  early chemical enrichment.  On the other hand, outflows remove metals from galaxies and tend to slow down the growth of their metal content at later time.  These ingredients are usually included in the simulation of dwarf galaxies, but with different levels of complexity and different numerical methods,  including one-zone models (e.g., \citealt{c02,lm03,lm04,fggl06,g07,lm10,lmc06,vmvl14,h15,k15,u15a}), semi-analytical models (e.g., \citealt{rs13,r15}), and hydrodynamical simulations (e.g., \citealt{k06,m08,r09,p12,rj12,hirai15,r16}).

Even in one-zone models, there are several ways to implement galactic inflows and outflows.  The goal of this paper is to determine whether the choice of implementation affects our ability to reproduce observed chemical evolution trends.  To do so, we target Sculptor, a dwarf spheroidal galaxy that is a satellite of the Milky Way.  Because it is a low-mass system (\citealt{b08,sfw10}), Sculptor must have been strongly affected by galactic outflows during its evolution (e.g., \citealt{s12,m15,afk16}), which offers a good opportunity to test our different outflow models. This dwarf galaxy has been simulated many times in the past (\citealt{lm03,lm04,fggl06,k06,g07,rs13,vmvl14,h15,k15,r16}) and therefore represents a testbed for our new chemical evolution model. In addition, the quality and quantity of observational data available for this galaxy (see Section~\ref{sect_sculptor_data}) allows us to better define the quality of our numerical predictions.

To conduct our experiment, we use three different implementation options for the circulation of gas and try to fit the observed stellar abundances of nine elements using OMEGA, our chemical evolution code.  To be rigorous in our comparisons, we combine OMEGA with the Markov Chain Monte Carlo (MCMC) code described in \cite{fhlg13} to derive the set of parameters for each model that best fits the observed stellar abundances in Sculptor.  All numerical predictions and observational data shown in this work are normalized to the solar composition found in \cite{gn93}, which is the one adopted in NuGrid models (see \citealt{p13}).

The use of the MCMC technique is based on a stochastic exploration of the parameter space, and offers several advantages for the analysis of our numerical framework.  As an example, in addition to determining the set of input parameters that best reproduces observations, an MCMC calculation provides the plausible range of values and the plausible correlations in the parameter space.  As a result, one can examine the consistency between different modeling assumptions and identify the degeneracies between parameters, which offer a complete and detailed view of the properties of a model.

This paper is organized as follows. We present in Section~2 the OMEGA single-zone chemical evolution model along with three prescriptions for the implementation of galactic inflows and outflows. In Section~3, we describe the general workflow of an MCMC calculation and explain the meaning of its outputs.  The results of the coupling between OMEGA, the MCMC code, and the stellar abundances observed in Sculptor are presented in Section~4.  We discuss our findings in Section~5 and then present our conclusions in Section~6.

\section{The OMEGA Code}
\label{sect_omega}
OMEGA (One-zone Model for the Evolution of GAlaxies) is a Python code designed to reproduce in a simple way the chemical evolution of local galaxies with known star formation histories (SFHs).  It is part of a numerical pipeline that aims to create connections between the areas of nuclear physics, stellar evolution, galaxy formation and evolution, and observations.  The closed-box version of OMEGA and our assumptions regarding the treatment of SSPs using SYGMA (Stellar Yields for Galactic Modeling Applications, Ritter et al.~in prep.) are described in detail in \cite{c15}.  In the following sections, we present the open-box version of the code along with three different implementations for the treatment of galactic inflows and outflows.  SYGMA, OMEGA, and STELLAB, our observational data plotting tool, are available online as part of the NuPyCEE\footnote{\href{http://github.com/NuGrid/NuPyCEE}{http://github.com/NuGrid/NuPyCEE}} (NuGrid Python Chemical Evolution Environment) package.

\subsection{The Main Equation}
\label{sect_tme}
Because OMEGA is a one-zone model, it is straightforward to describe its evolution.  At every time $t$, the mass of the gas reservoir, $M_\mathrm{gas}$, is updated to a new time $t+\Delta t$ by solving the equation
\begin{multline}
M_{\rm gas}(t+\Delta t)=M_{\rm gas}(t) +  \\
\Big[\dot{M}_{\rm in}(t) +\dot{M}_{\rm ej}(t) - \dot{M}_\star(t) - \dot{M}_{\rm out}(t)\Big]\Delta t,
\label{eq_main}
\end{multline}
where $\Delta t$ is the duration of the timestep.  From left to right, the four terms in brackets represent the inflow rate of gas into the galaxy ($\dot{M}_{\rm in}$), the rate at which stars release mass into the galaxy ($\dot{M}_{\rm ej}$), the star formation rate (SFR; $\dot{M}_\star$), and the rate at which gas flows out of the galaxy as a result of stellar feedback ($\dot{M}_{\rm out}$).  This equation is used in all of our three implementations of galactic inflow and outflow (see next sections).  As in \cite{c15}, the stellar ejecta rate includes the mass ejected from all SSPs that have been formed by time $t$, where the age, mass, and metallicity of each SSP is taken into account using SYGMA (Ritter et al.~in prep.).  The SFH is taken from observations (\citealt{db12} for Sculptor) and is an input to the model.  The unknowns of equation~(\ref{eq_main}) are the total mass of gas and the inflow and outflows rates.  The presence of gas stripping during the evolution of our galaxies is not included in equation~(\ref{eq_main}), but we refer to Section~\ref{sect_disc_add_phy_ing} for a discussion. We refer to \cite{afk16} for a more realistic view of the gas circulation processes throughout the evolution of galaxies.

The mass of each element ejected by stars is calculated using the stellar yields described in Section~\ref{sect_yields}.  We assume uniform mixing but we consider the different delay times between the formation of stars and the release of their ejecta.  The mass locked away by star formation and ejected by galactic outflows at time $t$ possess the chemical composition of the gas reservoir at that time.  For galactic inflows, we assume a primordial composition.

\subsection{Simple Inflow/Outflow Model}
\label{sect_sioc}
This model, which we hereafter refer to as the \textit{IO model} (for Inflow and Outflow), considers constant ratios between the star formation rate, galactic inflow, and galactic outflow throughout the evolution of the galaxy.  Assuming that galactic outflows are driven entirely by stellar activity, the outflow rate is derived from the SFR (e.g., \citealt{mqt05}),
\begin{equation}
\dot{M}_{\rm out}(t)=\eta\dot{M}_\star(t),
\label{eq_eta_gen}
\end{equation}
\noindent where $\eta$ is the mass-loading factor and is a free parameter.  This parametrization has been used many times in hydrodynamical simulations and in semi-analytical models of galaxy formation and evolution (see \citealt{sd14} and references therein).  Using equation~(\ref{eq_eta_gen}) alone, however, is not representative of galaxies substantially more massive than the Milky Way, since the gas dynamics in such high-mass systems is believed to be driven by active galactic nuclei feedback, which is not strongly coupled to the star formation rate (e.g., \citealt{f12}).

We recall that we take into account the delay between star formation and the release of stellar ejecta.  Therefore, to be more realistic, the galactic outflow rate should be proportional to the rate of mechanical energy injected by stars (e.g., \citealt{y13,cmd15}). Otherwise, when the outflow rate is proportional to the SFR, as in our model, a stellar population releases its ejecta after the occurrence of the galactic outflow triggered by that population. We tested a delayed-outflow prescription where the rate was proportional to $\dot{M}_\mathrm{ej}$ but found no major difference in our results. This is mainly because our one-zone model does not calculate the interplay between stellar feedback, star formation, and gas flows between different gas components (e.g., between the cold and hot phases in semi-analytical models). More complex galaxy models that self generate the SFH are sensitive to different outflow prescriptions (see \citealt{y13}). However, delayed outflows could also alter the chemical evolution predicted by OMEGA when considering episodic SFHs with short periods or SFHs that decline on a shorter timescale than in the case of Sculptor.

In the IO model, we impose a proportionality between the rates of gas inflow and outflow in the galaxy,
\begin{equation}
\dot{M}_{\rm in}(t)=\xi\dot{M}_{\rm out}(t).
\label{eq_in_out_rel}
\end{equation}
where $\xi$ is a free parameter used to regulate the relative intensity of inflowing material.  In theory, the star formation rate should be proportional to the inflow rate and the latter should scale with the total mass of the galaxy, including its dark matter halo.  However, in order to easily mimic the evolution of observed galaxies, the SFH is an input to our model and is not calculated from the inflow rate.  The relation between inflows and outflows in the IO model (equation~\ref{eq_in_out_rel}) is therefore a way to link the star formation rate to the inflow rate, as the outflow rate is directly proportional to the star formation rate (see equation~\ref{eq_eta_gen}).  Furthermore, the total mass of the galaxy is not included in the equations of the IO model (but see Section~\ref{sect_mah}).

Once the initial mass of gas at $t=0$ is set manually, equation (\ref{eq_main}) can be used to calculate the evolution of the gas reservoir and chemical abundances at every timestep.  We note that this model represents a simplification, since the star formation process also depends on the physical conditions of the interstellar medium (e.g., \citealt{mo07,stl14}), which we do not consider. We refer to Section~\ref{sect_disc_add_phy_ing} for a discussion on neglected physical processes.

\subsection{Star Formation Model}
\label{sect_kssfl}
Star formation in galaxies is tightly correlated to the density of the interstellar medium (\citealt{s59,k98}).  This so called Kennicutt-Schmidt law is used in almost every semi-analytical model of galaxy evolution, in the adapted form of(\citealt{kauf93,kcdw99,cole94,clbf00,swtk01,baugh06,sd14})
\begin{equation}
\label{eq_SF_law}
\dot{M}_\star(t)=\epsilon_\star\frac{M_\mathrm{gas}(t)}{\tau_\star},
\end{equation}
where $\epsilon_\star$ and $\tau_\star$ are  the star formation efficiency and the star formation timescale, respectively.  In this model, hereafter referred to as the \textit{SF model} (for Star Formation), $\epsilon_\star$ and $\tau_\star$ are both constant quantities and are thus merged into a single constant free parameter,
\begin{equation}
f_\star = \frac{\epsilon_\star}{\tau_\star}.
\end{equation}
Since $\dot{M}_\star$ is known at every timestep, equation (\ref{eq_SF_law}) can then be inverted and used to calculate the mass of gas at any time $t$.  In this model, we still use equation~(\ref{eq_eta_gen}) to set the outflow rate.

Since the mass of gas is now a known quantity, equation~(\ref{eq_main}) is rearranged in order to solve for the inflow rate at each timestep.  This approach has also been used by others to calculate the chemical evolution of local dwarf spheroidal galaxies (\citealt{fggl06,g07,h15}), since it ensures a coherent connection between the SFR and all the different terms found in equation~(\ref{eq_main}). In the case of Sculptor, the star formation rate eventually drops to zero after several Gyr of evolution (\citealt{db12}). With the SF model, and also with our third model described in Section~\ref{sect_mah}, this implies that the mass of gas will also drop to zero at the end of our simulations. During the last timesteps, if the inflow rate becomes negative in order to empty the gas reservoir, the inflow rate is set to zero and the outflow rate is momentarily increased (see \citealt{h15}).  Although this gas removal is rather artificial, it could in principle be associated with a gas stripping process (see Section~\ref{sect_dbm_disc}).

\subsection{Mass Assembly Model}
\label{sect_mah}
In the $\Lambda$CDM hierarchical scenario, low-mass dark matter halos form at high redshifts and progressively increase their mass via mergers (\citealt{b84,mw02,cf05}).  The stellar component of galaxies are typically found at the center of each galaxy's dark matter halo.  Although baryons and dark matter behave differently, galaxies and dark matter halos should have similar merger histories, in terms of how many mergers occur as a function of time.  Within this framework, galactic chemical evolution in an open box is not only about including gas inflows and outflows as in the IO and SF models, but is also about considering the time evolution of the total mass of the system.  The mass of the dark matter halo has a significant impact on the evolution of galaxies (e.g., \citealt{behroozi13,moster13,munshi13}).  In particular, compared to Milky Way-size galaxies, low-mass galaxies with their shallow gravitational potential wells are more vulnerable to stellar feedback (e.g., \citealt{maclow99,stinson09,hopkins14}).  Our last model, hereafter referred to as the \textit{MA model} (for Mass Assembly), is an extension of the SF model where we include a simplified version of the hierarchical scenario in order to account for the impact of dark matter on galactic outflows and star formation timescales.

\subsubsection{Mass-Dependent Mass-Loading Factor}
When a simulated galaxy grows significantly during its lifetime, it becomes important to add a mass dependency to the mass-loading factor $\eta$.  Although the typical observed values for this parameter usually range between 0.01 and 10 (\citealt{vcb05}), simulations suggest that $\eta$ can reach values up to $\sim100$ in the case of dwarf galaxies (e.g. \citealt{s12,m15}).  In this model, equation~(\ref{eq_eta_gen}) is still used to calculate the galactic outflow rate, but the constant $\eta$ associated with the IO model is substituted by the redshift- and mass-dependent $\eta$ defined in this section.  Following the development of \cite{mqt05}, the mass-loading factor can be defined as
\begin{equation}
\label{eq_ml_v}
\eta\propto v_\mathrm{out}^{-\gamma},
\end{equation}
where $v_\mathrm{out}$ is the velocity of the outflowing material.  According to \cite{mqt05}, $\gamma=1$ or 2 when outflows are either driven by the transfer of momentum or by the energy emerging from stellar activity.  However, in our model, we consider $\gamma$ as a free parameter that we call the mass-loading power-law index.

Observations and simulations have shown that $v_\mathrm{out}$ is proportional to the rotation velocity of galaxies (\citealt{martin05}) and therefore to the circular velocity $V_\mathrm{vir}$ of the host virialized systems (\citealt{m15}).  This last velocity is defined by the Virial theorem,
\begin{equation}
\label{eq_v_vir}
V_\mathrm{vir}^2=\frac{GM_\mathrm{vir}}{R_\mathrm{vir}},
\end{equation} 
where $G$, $M_\mathrm{vir}$, and $R_\mathrm{vir}$ are the Newton gravitational constant and the mass and radius of the virialized system, respectively.  Following \cite{wf91}, the virial radius can be defined by
\begin{equation}
\label{eq_R_vir}
R_\mathrm{vir}=0.1H_0^{-1}(1+z)^{-3/2}V_\mathrm{vir},
\end{equation}
where $H_0$ and $z$ are the present-day value of the Hubble parameter and the redshift, respectively.  By substituting this last relation in equation~(\ref{eq_v_vir}) and by solving for $V_\mathrm{vir}$, the mass-loading factor defined in equation~(\ref{eq_ml_v}) can be rewritten as
\begin{equation}
\label{eq_eta_mdep}
\eta(z)=C_\eta M_\mathrm{vir}^{-\gamma /3}(1+z)^{-\gamma/2},
\end{equation}
where $M_\mathrm{vir}$ also varies with redshift (see Section~\ref{sect_aaah}). Redshift is converted into time using the WMAP5 cosmological parameters (\citealt{d09}).  The normalization constant $C_\eta$ is calculated by
\begin{equation}
C_\eta=\eta(z=0)M_\mathrm{vir}^{\gamma /3},
\end{equation}
where $\eta(z=0)$,  the value of $\eta$ at the end of a simulation, is the actual free parameter regulating the strength of the outflows.  We refer to \cite{hqm12} and \cite{m15} for alternate mass-loading factor relations, which are derived from hydrodynamic simulations.

\begin{deluxetable*}{ccccc}
\tabletypesize{\footnotesize}
\tablewidth{0pc}
\tablecaption{Best Values of the Free Parameters, Along with their 68\,\% Confidence Intervals, used in our Models to Fit the Chemical Evolution of Sculptor\label{tab_free_param}}
\tablehead{ \colhead{\multirow{2}{*}{Parameter}} & \colhead{\multirow{2}{*}{Description}} & \colhead{} & \colhead{Best value} & \colhead{} \\
  \colhead{} & \colhead{} & \colhead{IO model} & \colhead{MA model} & \colhead{SF model} }
\startdata
$N_{\rm Ia}$ & Number of SNe Ia [$10^{-3}$ M$_\odot^{-1}$] & $\mathbf{2.6^{+0.6}_{-0.3}}$ & $\mathbf{1.6^{+0.2}_{-0.2}}$ & $\mathbf{1.0^{+0.2}_{-0.1}}$ \\
\noalign{\medskip}
$M_{\rm trans}$ & Minimum stellar initial mass for CC~SNe [M$_\odot$] & $\mathbf{11.0^{+0.6}_{-0.6}}$ & $\mathbf{10.3^{+0.7}_{-0.7}}$ & $\mathbf{10.9^{+0.5}_{-1.2}}$ \\
\noalign{\medskip}
$\eta$ & Mass-loading factor (ratio between $\dot{M}_\mathrm{out}$ and $\dot{M}_\star$) & $\mathbf{6.0^{+3.4}_{-2.0}}$ & $^a$$\mathbf{41.0^{+8.1}_{-7.0}}$ & $\mathbf{15.7^{+2.5}_{-1.0}}$  \\
\noalign{\medskip}
$\xi$ & Ratio between the inflow and outflow rates & $\mathbf{1.70^{+1.3}_{-0.3}}$ & $<1.1>$ & $<1.0>$  \\
\noalign{\medskip}
$M_{\rm gas}$ & Initial mass of the gas reservoir [$10^6$ M$_\odot$] & $\mathbf{8.0^{+3.3}_{-2.1}}$ & 1.4 & 3.4 \\
\noalign{\medskip}
$f_\star$ & Star formation efficiency [10$^{-10}$ yr$^{-1}$] & $<0.37>$ & $<4.8>$ & $\mathbf{6.1^{+13.2}_{-1.3}}$ \\
\noalign{\medskip}
$R_{\star,\mathrm{dyn}}$ & Ratio between $\epsilon_\star$ and $f_\mathrm{dyn}$ & --- & $\mathbf{0.095^{+0.18}_{-0.03}}$ & ---  \\
\noalign{\medskip}
$\gamma$ & Mass-loading power-law index & --- & $\mathbf{3.3^{+0.5}_{-0.8}}$ & --- \\
\noalign{\medskip}
\enddata
\tablecomments{Values in boldface are associated with the parameters included in the MCMC calculation, where the best values correspond to the peak values of the resulting probability distribution functions.  The values in normal character were calculated by the model and extracted, when possible, from a simulation ran with the best set of parameters.  Values in brackets represent an average over the entire active star formation period.  All models are described in Section~\ref{sect_omega}.}
\tablenotetext{a}{This is only the final value at the end of the simulations.  The average value is 26.8.}
\end{deluxetable*}

\subsubsection{Averaged Accretion History}
\label{sect_aaah}
As opposed to old stars, which track the SFH of galaxies, there is no clear observational evidence to directly constrain the evolution of the gas content of a specific galaxy as a function of its age.  It is even more difficult to derive its dark matter mass assembly, which is nevertheless an essential part of setting the efficiency of galactic outflows.  The evolution of dark matter halos can be captured by large-scale numerical simulations, such as Millennium~I and II (\citealt{s05,bk09}), Bolshoi (\citealt{ktp11}), the Illustris (\citealt{v14}) and EAGLE (\citealt{schaye15}) projects\footnote{The Illustris and EAGLE simulations also include baryons.}, and the Caterpillar project (\citealt{g15}).  In this work, we use the average accretion rate extracted from the Millennium~II simulation by \cite{fak10} to follow the evolution of the mass of dark matter, $M_\mathrm{DM}$, as a function of time,
\begin{multline}
\dot{M}_{\rm DM}=46.1\left(\frac{M_{\rm DM}}{10^{12}\,{\rm M}_\odot}\right)^{1.1}\\
(1+1.11z)\sqrt{(\Omega_m(1+z)^3+\lambda}\quad[{\rm M}_\odot\,\rm yr^{-1}]\,.
\label{eq_dmdt_dm}
\end{multline}
The values for the mass and dark energy densities, $\Omega_m$ and $\lambda$, are taken from \cite{d09}.

We integrated equation~(\ref{eq_dmdt_dm}) several times with different initial $M_\mathrm{DM}$ values at $z=14$ and built a database of pre-calculated mass evolution paths for dark matter halos.  In the MA model, the key parameter defining the evolution of $M_\mathrm{DM}$ is the dark matter mass at $z=0$.  From this parameter, the code interpolates the database to select the appropriate averaged mass evolution path.  As a result, the complete mass evolution of dark matter as a function of time is known before running an OMEGA simulation.  For the current dark matter mass of Sculptor we use the estimated mass provided by \cite{sfw10}. Throughout this paper, we assume that $M_\mathrm{DM}\sim M_\mathrm{vir}$ since dwarf galaxies are typically dominated by dark matter (e.g., \citealt{behroozi13,moster13,munshi13}).  We therefore substituted $M_\mathrm{vir}$ by $M_\mathrm{DM}$ in equation~(\ref{eq_eta_mdep}) in all of our simulations.

\subsubsection{Time-Dependent Star Formation Timescale}
\label{sect_tdsft}
As in several semi-analytical models, we assume that $\tau_\star$, the star formation timescale, is proportional to the dynamical timescale, $\tau_\mathrm{dyn}$, of the whole virialized system (e.g. \citealt{kcdw99,clbf00,swtk01}), which includes dark matter and baryons.  With $\tau_\mathrm{dyn}\approx R_\mathrm{vir}/V_\mathrm{vir}$, which can be extracted from equation~(\ref{eq_R_vir}), the star formation timescale can be defined by
\begin{equation}
\label{eq_tau_z}
\tau_\star=f_\mathrm{dyn}\tau_\mathrm{dyn}\approx0.1f_\mathrm{dyn}H_0^{-1}(1+z)^{-3/2}.
\end{equation}
In the MA model, equation~(\ref{eq_SF_law}) is still used to calculate the mass of gas at each timestep, but the constant $\tau_\star$ associated with the IO model is substituted by the redshift-dependent $\tau_\star$ defined in equation~(\ref{eq_tau_z}).  Since $\epsilon_\star$, used in equation~(\ref{eq_SF_law}), and $f_\mathrm{dyn}$ are both constant quantities, we merged them into a single parameter defined by
\begin{equation}
R_{\star,\mathrm{dyn}} = \frac{\epsilon_\star}{f_\mathrm{dyn}}.
\end{equation}
With this prescription, the mass of the gas reservoir will generally tend to increase with time, unless the star formation rate drops to zero, in which case the mass of gas will also drop to zero.

\subsection{Stellar Yields}
\label{sect_yields}
As the foundation of our chemical evolution simulations, we used the stellar yields calculated by the NuGrid collaboration for low- and intermediate-mass stars and for massive stars (Ritter et al. in prep.).  This consistent set of yields includes five initial metallicities from $Z=10^{-4}$ to $0.02$ in mass fraction and 12 stellar models per metallicity with initial masses from 1 to 25~M$_\odot$.  For this study, we used the yields associated with version 1.0 of the online NuPyCEE package (see Section~\ref{sect_omega} for links) and chose the set of core-collapse supernova (CC~SN) yields that has been calculated with a mass-cut prescription based on the electronic fraction ($Y_e$) in the stellar interior.  The mass-cut represents the location inside a massive star where the explosion is launched.  For all explosive yields, we set the mass-cut to the location where $Y_e=0.4992$.  We refer to \cite{p13} and Ritter et al. (in prep.) for explosive yields with an alternative mass-cut prescription.

The minimum mass for CC~SNe ($M_\mathrm{trans}$), setting the transition between intermediate-mass and massive stars, is left as a free parameter for the MCMC calculations and is allowed to range from 8 to 12\,M$_\odot$ (see Section~\ref{sect_mmCCSNe}). Throughout this paper, we assume that stars more massive than 30\,M$_\odot$ produce black holes and do not contribute to the stellar ejecta (see Section 3.9.1 in \citealt{c15} for a discussion). The stellar models at $Z=10^{-4}$, the lowest metallicity provided by NuGrid, are used at the beginning of each simulation until the metallicity of the gas reservoir reaches $Z=10^{-4}$.  Beyond this point, stellar yields are interpolated as a function of metallicity.

\begin{figure*}
\includegraphics[width=7in]{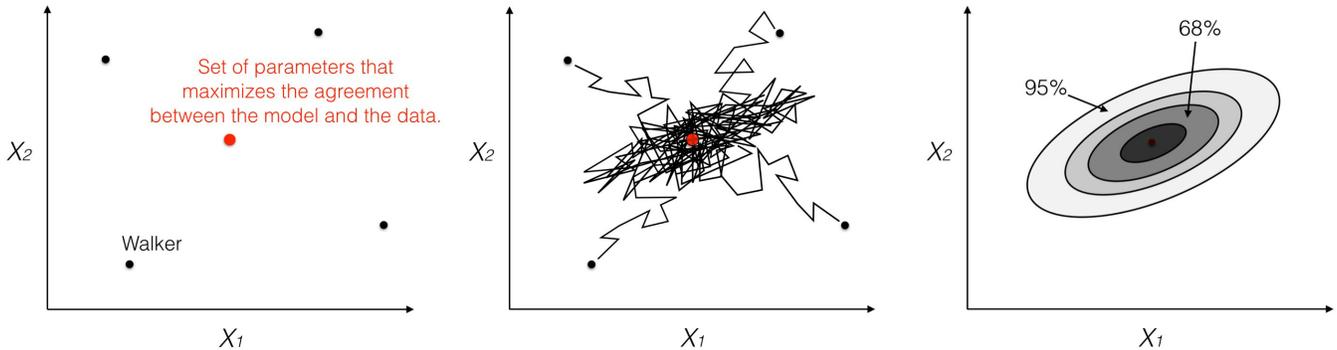}
\caption{Diagram of how the Markov Chain Monte Carlo algorithm works for a model with two parameters, $X_1$ and $X_2$.  Left panel: Walkers are distributed randomly in the allowed parameter space.  The red dot corresponds to the solution that provides the best fit between the model and the data, which is unknown prior the calculation.  Middle panel: Walkers explore the parameter space to find the best solution.  Their displacement is based on probability and random numbers, and controlled by the relative goodness of the fits associated with the walker's current and potential future positions.  Right panel: Contours are drawn to quantify how frequently walkers traversed each portion of the parameter space.}
\label{fig_MCMC_schema}
\end{figure*}

The ejecta of Type~Ia supernovae (SNe~Ia) has been generated by combining the yields of \cite{t86} to a power-law delay-time distribution (DTD) function in the form of $t^{-1}$.  This choice of DTD function is motivated by several observational studies (see \citealt{mmn14} and Table~4 in \citealt{c15}) and by the higher SNe~Ia rates observed in bluer galaxies (\citealt{mannucci05,li11}), which implies a prompt appearance of SNe~Ia in stellar populations.  We refer to \cite{c15} for more information about our SNe~Ia implementation and for references regarding alternative DTD functions. The normalization of the DTD function, the total number of SNe~Ia per stellar mass formed ($N_\mathrm{Ia}$), is left as a free parameters for the MCMC calculations and is allowed to range from 0 to $6\times10^{-3}$~$M^{-1}_\odot$.

\section{Markov Chain Monte Carlo}
MCMC methods are a class of algorithms that sample from a probability distribution to generate a random walk in the parameter space of a model.  This is often used to find the parameters of a model that best fits a certain collection of data (\citealt{m53,h70,grs98}).  In our case, the model is OMEGA and the data are the stellar abundances observed in Sculptor (see Section~\ref{sect_results}).  With an MCMC calculation, the best set of parameters is given in terms of probability distribution functions (PDFs) that can be used to quantify the likely values of each parameter.  This approach has already been used in different areas of astrophysics, such as cosmology (\citealt{d05}), cosmic rays (\citealt{pdm10}), active galactic nuclei (\citealt{r12}), Milky Way satellites (\citealt{u15b}), and semi-analytical models of galaxy formation (\citealt{kam08,g14,htor09,henri13}).  For the present paper, we used the publicly-available MCMC code \textit{emcee} (\citealt{gw10,fhlg13})\footnote{\href{http://dan.iel.fm/emcee}{http://dan.iel.fm/emcee}, \href{https://github.com/dfm/emcee}{https://github.com/dfm/emcee}} for maximum likelihood estimation.

\begin{figure*}
\includegraphics[width=7in]{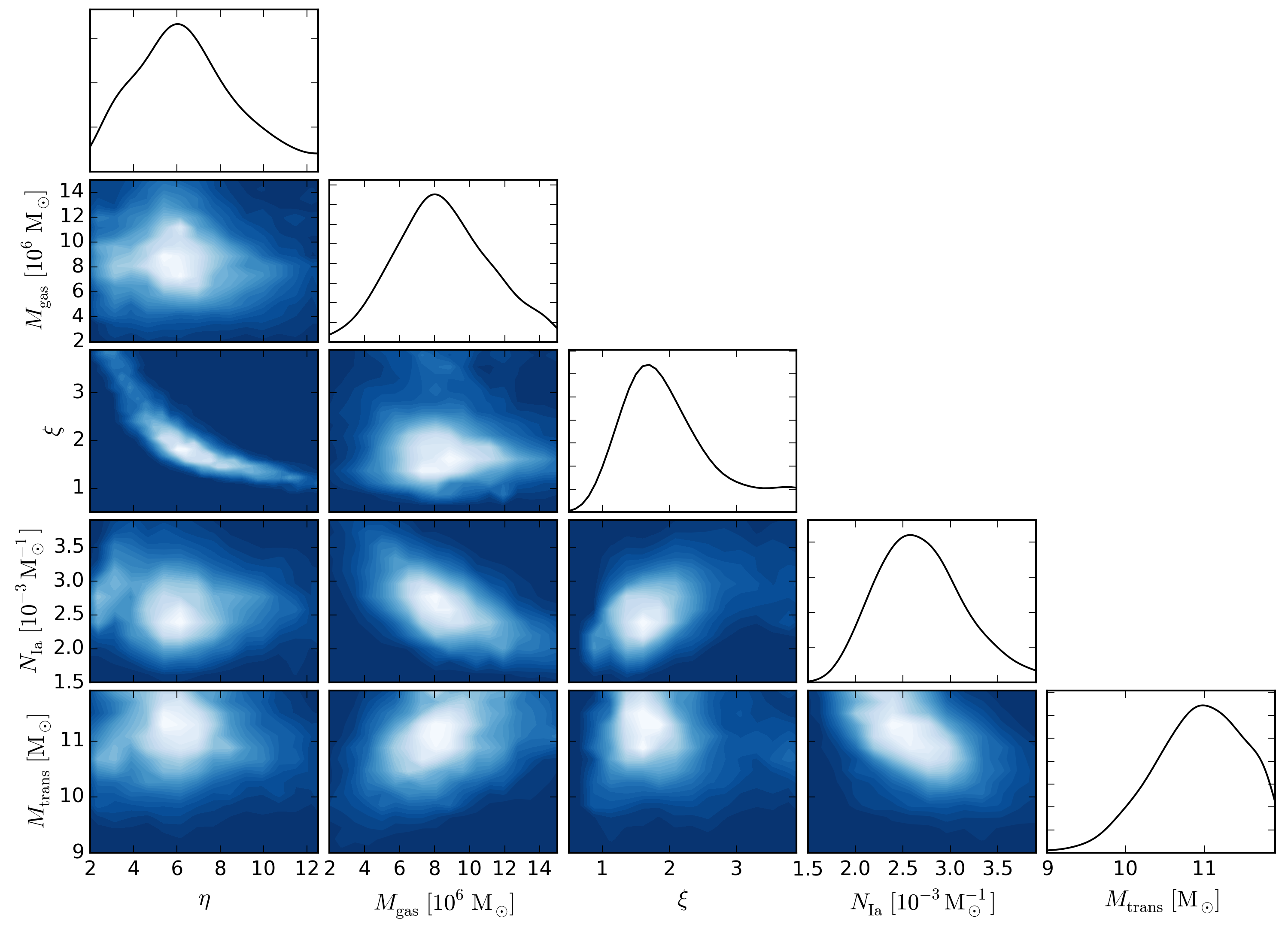}
\caption{Probability distribution of the input parameters of the IO model that provides the best fit with the chemical evolution of Sculptor.  The upper panel of each column shows the projected distribution of each parameter, where the Y axis is in arbitrary units.  The parameters are the mass-loading factor ($\eta$), the initial mass of gas ($M_\mathrm{gas}$), the ratio between inflow and outflow rates ($\xi$), the number of SNe~Ia per stellar mass formed ($N_\mathrm{Ia}$), and the minimum mass for CC~SNe ($M_\mathrm{trans}$).}
\label{fig_IO}
\end{figure*}

\subsection{General Workflow of an MCMC Calculation}
Figure \ref{fig_MCMC_schema} illustrates the main steps of an MCMC calculation.  First, a certain number of coordinates are chosen randomly within the parameter space of the model (left panel).  Those initial coordinates are called \textit{walkers} as they all represent the starting point ($\Theta_0$) of an exploratory itinerary, or a series of displacements inside the parameter space, that will be followed step by step by the MCMC code.  At each step, for each walker, a new set of parameters ($\Theta_{\mathrm{new}}$) is obtained by generating random perturbations around the walker's current position ($\Theta_{\mathrm{cur}}$).  The question then is whether the walker will remain at its current location or move to the new set of parameters.  This decision is taken by calculating a probability that defines the goodness of $\Theta_\mathrm{new}$ in fitting the data, which is done with a likelihood\footnote{The likelihood function adopted in this work as well as an explanation of how it is used for maximum likelihood estimation in the \textit{emcee} code can be found at \href{http://dan.iel.fm/emcee/current/user/line/\#maximum-likelihood-estimation}{http://dan.iel.fm/emcee/current/user/line/\#maximum-likelihood-estimation}.} function that quantifies the offsets between the predictions and the data.  If $\Theta_\mathrm{new}$ produces a better fit than $\Theta_\mathrm{cur}$, there will be a high probability for the walker to take the step and move to the new coordinate.  But if the new fit is worse, the probability will be low (but non-zero).  Once the probability is known, a random number is generated and associated with that probability to determine if the walker moves to the new location or stays at its current location (see \citealt{fhlg13} for more details).  Those operations are repeated over and over until the optimal solution is found (middle panel of Figure~\ref{fig_MCMC_schema}).

Because the trajectories are calculated in a probabilistic manner, the end point of the walkers, or their final destination, is not meant to be the best solution.  In other words, once the walkers isolate the best area in the parameter space, they start to explore the different nearby solutions.  However, since the walkers are attracted toward the best set of parameters (e.g., the red dot in Figure~\ref{fig_MCMC_schema}), they will still mostly be found near that specific solution.  Once an MCMC calculation is over, the trajectory of all walkers are combined and transformed into a density map, such as the one shown in the right panel of Figure~\ref{fig_MCMC_schema}.  This is done by calculating how many times walkers have been found within a certain area of the parameter space.  For example, the 95\% contour means that the walkers have spent 95\% of their time, or steps, within the considered area.  These percentages actually represent the confidence levels or the probability of having found the best set of parameters.  In addition, the shape on the contours highlights the degeneracy between the parameters, as will be shown in Section~\ref{sect_results}.

\subsection{Convergence}
The number of walkers and the number of steps considered in an MCMC calculation can have a significant impact on the results (see \citealt{fhlg13}).  Walkers can be trapped in local maxima and miss the optimal solution.  Therefore, if not enough walkers are used, the solutions found by the MCMC calculation may not be a good representation of the entire parameter space.  Also, if not enough steps are followed, the walkers may not have enough \textit{time} to find the best set of parameters.  In all of the simulations presented in this work, we made several tests to ensure we have reached convergence, in the sense that the best fit recovered by the MCMC calculations is not significantly modified by considering more walkers or more steps.

\section{Application to Sculptor}
\label{sect_results}
In this section we present the results of our MCMC calculations for each of the models described in Section~\ref{sect_omega}.  We considered 480 walkers and followed them for 900 steps, corresponding to more than 400,000 chemical evolution simulations per model.  The list of parameters included in the MCMC calculations along with their best values are presented in Table~\ref{tab_free_param}.  In addition to the parameters described in Section~\ref{sect_omega}, we also included the number of SNe~Ia per stellar mass formed ($N_\mathrm{Ia}$) and the minimum mass for CC~SNe ($M_\mathrm{trans}$), which are common is all three models.

\begin{figure*}
\includegraphics[width=7in]{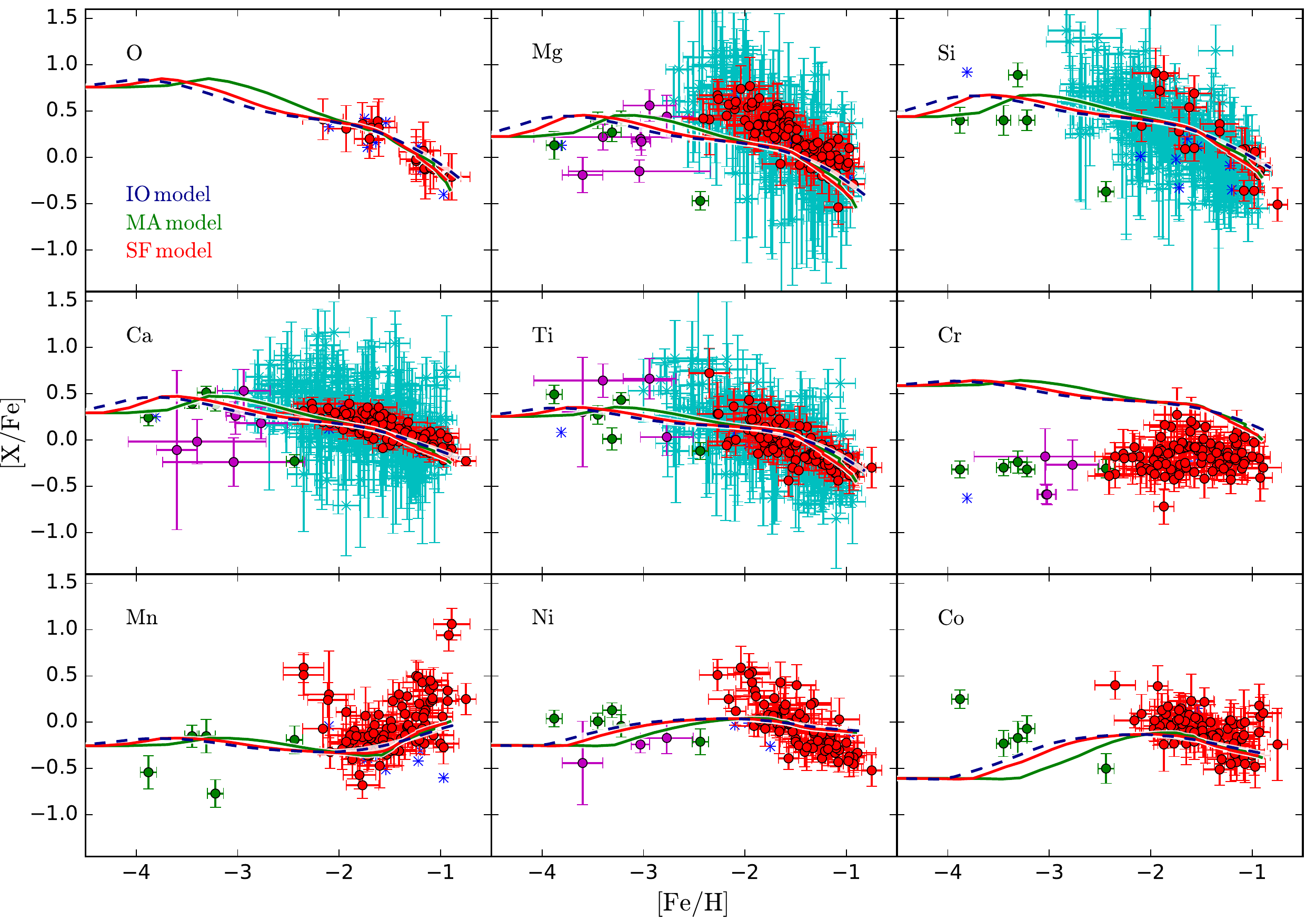}
\caption{Best fit for the chemical evolution of Sculptor using the IO (blue), MA (green), and SF (red) models, for nine elements.  Observational data come from Frebel et al.\,(\citeyear{f10}, blue crosses), Kirby et al.\,(\citeyear{k10}, cyan crosses), Starkenburg et al.\,(\citeyear{s13}, purple circles), Jablonka et al.\,(\citeyear{j15}, green circles), and Hill et al.\,(priv. comm., red circles).}
\label{fig_fit_IO}
\end{figure*}

During the process of quantifying goodness of fit between the data and a given model, more weight has been paid to the data points that have the smallest error bars and to the regions in the [X/Fe] vs [Fe/H] space that have the highest concentration of data points (e.g., [Fe/H]~$\gtrsim-2.5$).  The overall statistical weight of each element is defined by both the number of data points and their precision.  This work represents the first exploratory steps of coupling our chemical evolution tools with an MCMC code, which is why we only included nine elements (O, Mg, Si, Ca, Ti, Cr, Mn, Ni, and Co) for the observational constraints in our MCMC calculations and only focused on one galaxy.
\\
\subsection{Sculptor Data}
\label{sect_sculptor_data}
There are now several detailed chemical analyses from high quality spectra of red-giant-branch stars 
in Sculptor (\citealt{s03,f10,t10,s13,j15,sk15a,sk15b}; Hill et al.\,in prep.)
Most of these analyses have been done with high signal-to-noise ratios ($>$~50) and with high resolution (R~$>$~45 000) VLT-UVES 
spectroscopy, though some are from lower resolutions (Magellan-MIKE spectra with R~$\sim$~32 000,
VLT-FLAMES-GIRAFFE spectra with R~$\sim$~20 000, and VLT-Xshooter spectra with R~$=$~11 000).
Regardless of the source, these analyses paint a similar picture of an early chemical evolution 
dominated by massive stars in a well mixed interstellar medium.  \cite{j15} suggest 
that the early chemical evolution of Sculptor was similar to that of other classical dwarf galaxies 
and the Milky Way halo, at least for the majority of old stars (80\,\%).

To mimic the evolution of Sculptor with OMEGA, we used the SFH provided by \cite{db12} as an input.  The estimated total (dark matter + baryons) mass of $1.5\times10^9$\,M$_\odot$ for this galaxy, which is needed for the MA model, has been taken from \cite{sfw10}. This is higher than the enclosed dynamical mass of $\sim3\times10^8$\,M$_\odot$ derived observationally by \cite{b08}.  However, in order to be consistent with the relation used in our model for the evolution of dark matter (see equation~\ref{eq_dmdt_dm}), we decided to use the mass derived by \cite{sfw10} to consider the entire dark matter halo, which should extend beyond the maximum observable radius defined by the stars.  As described in Section~\ref{sect_mah} for the MA model, the dark matter halo mass is only used to introduce a time dependence to the mass-loading factor.  The exact value of the dark matter content of Sculptor is therefore not so important, since the evolution of the mass-loading factor as a function of time is scaled by the input parameter $\eta$.

\begin{figure}
\begin{center}
\includegraphics[width=3.3in]{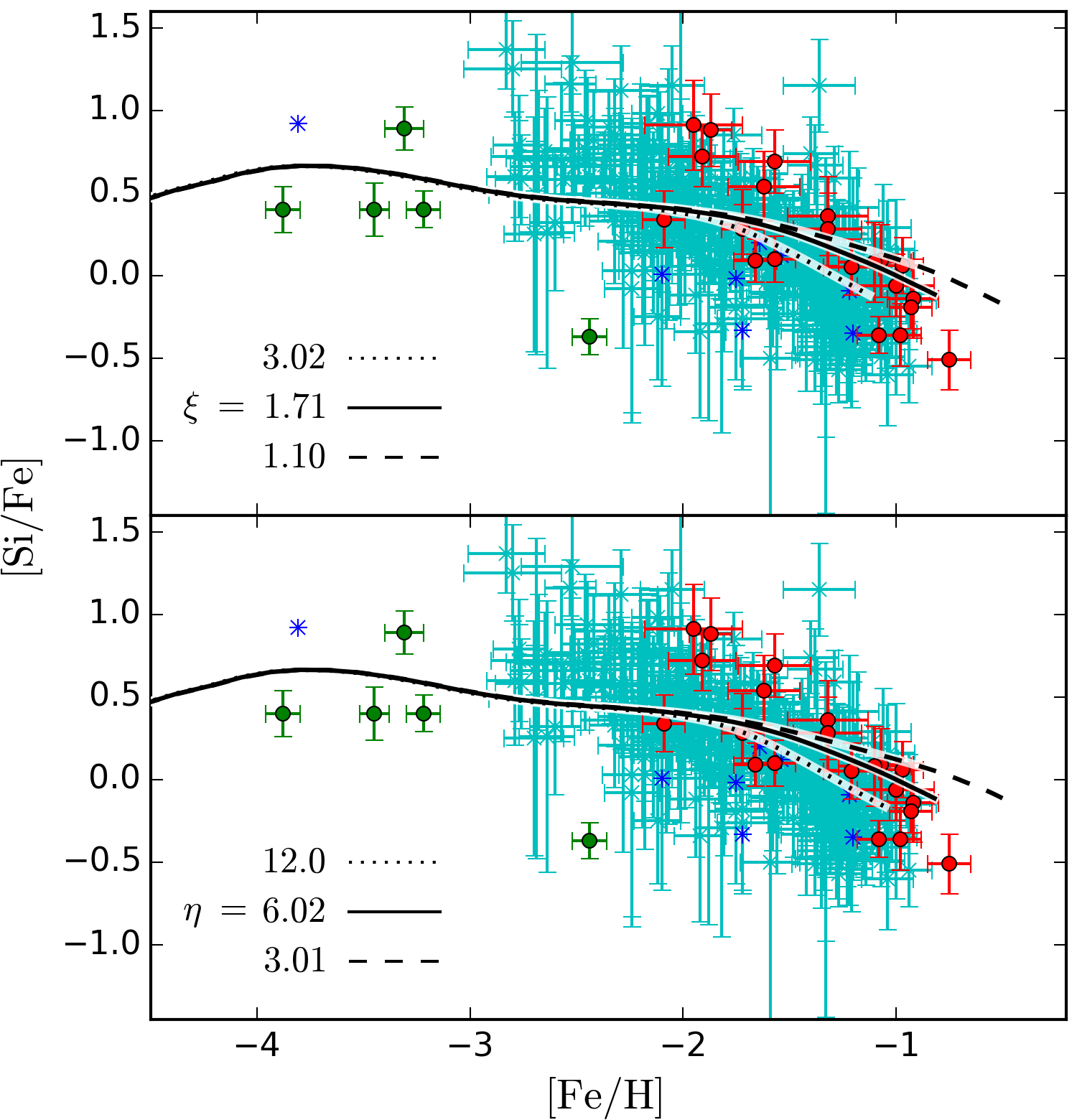}
\caption{Impact of inflows (upper panel) and outflows (lower panel) on the predicted evolution of Si as a function of [Fe/H] for the IO model.  The best fit resulting from the MCMC calculation is given by the solid line.  The observational data are the same as in Figure \ref{fig_fit_IO}.  The rate ratio between inflows and outflows ($\xi$) is limited to a minimum value of $\sim1.1$, below which not enough gas is introduced in the system to sustain the star formation history.}
\label{fig_in_out_deg}
\end{center}
\end{figure}

\subsection{The IO Model Results}
\label{sect_results_IO}
Figure \ref{fig_IO} presents the results of the MCMC calculation with the IO model, where each panel shows the probability distribution of a pair of two input parameters.  The values located within the brightest regions represent the most probable solution for fitting the stellar abundances observed in Sculptor. The histogram on top of each column is a projection of the parameter labeled at the bottom of that column, from which the confidence interval of the best solution is extracted.  A narrow distribution implies that the MCMC calculation efficiently constrained the considered parameter.  On the other hand, if the distribution is relatively broad, the best solution is not unique and a reasonable fit can be achieved by using a range of values (see also \citealt{h15}).

The blue lines in Figure~\ref{fig_fit_IO} present the best fit that we can recover with the IO model, given our choice of stellar yields.  The bump seen in our alpha element predictions at [Fe/H]~$<-3$ is caused by the ejecta of our 20\,M$_\odot$ stellar model associated with the lowest metallicity provided by NuGrid ($Z=10^{-4}$). This feature is visible with all three galaxy models since they use the same set of stellar yields.  Because of the relatively high remnant mass of the 20\,M$_\odot$ model compared to the adjacent 15 and 25\,M$_\odot$ models, most of the iron located in its core is not ejected, which leads to an ejecta with a higher [$\alpha$/Fe] ratio. The bump therefore represents the interpolated transitions between the 25, 20, and 15\,M$_\odot$ models. At [Fe/H]~$<-3$, the composition of the gas reservoir still shows the signature of individual massive star models, while at later times the gas reservoir represents a mixture of ejecta produced by many different stellar populations. The shape of numerical predictions highly depends on stellar yields (e.g., \citealt{rktm10,mcgg15}).  Features such as the bumps seen in Figure~\ref{fig_fit_IO} could completely disappear if different stellar models were selected in our set of yields (see \citealt{c16}).

\subsubsection{Degeneracy Between Inflows and Outflows}
As seen in Figure~\ref{fig_IO}, there are correlations between some of the parameters.  The degeneracy between the strength of outflows ($\eta$) and strength of inflows ($\xi$) is caused by two constraints. First, the walkers search for a solution where most of the observed stars are covered by the prediction.  Second, the walkers avoid solutions that predict the formation of stars at [Fe/H] $\gtrsim-1$, since such high-metallicity stars are not observed in Sculptor.  In other words, the MCMC calculation favours solutions where numerical predictions have a maximum [Fe/H] value similar to observations.  Although inflows and outflows are two different processes in terms of galaxy evolution, they both modify the high-metallicity end of our predictions in a similar way (see Figure~\ref{fig_in_out_deg}).  Galactic outflows remove iron from the galaxy and thus reduce the maximum [Fe/H] value (see also \citealt{andrews16}), whereas primordial gas inflows add hydrogen to the galaxy and dilute the metal content of the gas, which, if inflows occur before the end of the star-forming period, also reduces the maximum [Fe/H] value.  There are then several possible combinations that can be used to adjust the final [Fe/H] value in our simulations, thus creating the degeneracy between $\eta$ and $\xi$.

As seen in Figure~\ref{fig_in_out_deg}, modifying the rates of inflows and outflows simply results in stretching or shrinking the high-metallicity end of numerical predictions along the [Fe/H] axis.  As a matter of fact, in our model, these processes do not significantly modify the metal composition of the gas reservoir.  Even if outflows eject metals from the galaxy, the abundance ratios remain the same because of our uniform mixing assumption.  In the case of inflows with primordial composition, only hydrogen and helium are added to the system, which also do not modify the relative composition of the metals.

\subsubsection{Degeneracy Between Models}
\label{sect_res_deg_bet_mod}
In addition to the degeneracy seen between the different input parameters, there is also a degeneracy between our models (see Section~\ref{sect_dbm_disc}), in the sense that a similar answer can be recovered by either the IO, the MA, or the SF model (see Figure~\ref{fig_fit_IO}).  This is consistent with the work of \cite{rs13}, who used the Munich semi-analytical model on top of the Aquarius large-scale dark matter simulations (\citealt{springel08a,springel08b}).  They considered four Sculptor-like galaxies in a cosmological framework and compared their predicted chemical evolution with a model similar to OMEGA in terms of relative complexity.  Although the different mass assembly and SFHs associated with the four Sculptor-like galaxies generated significant scatter in the predictions, their global chemical evolution trends were all similar and consistent with the ones generated by their simple model.  Our results therefore reinforce the original findings of \cite{rs13}, who showed that different galaxy evolution models have limited impact on the chemical evolution of dwarf systems such as Sculptor.

\begin{figure*}
\includegraphics[width=7in]{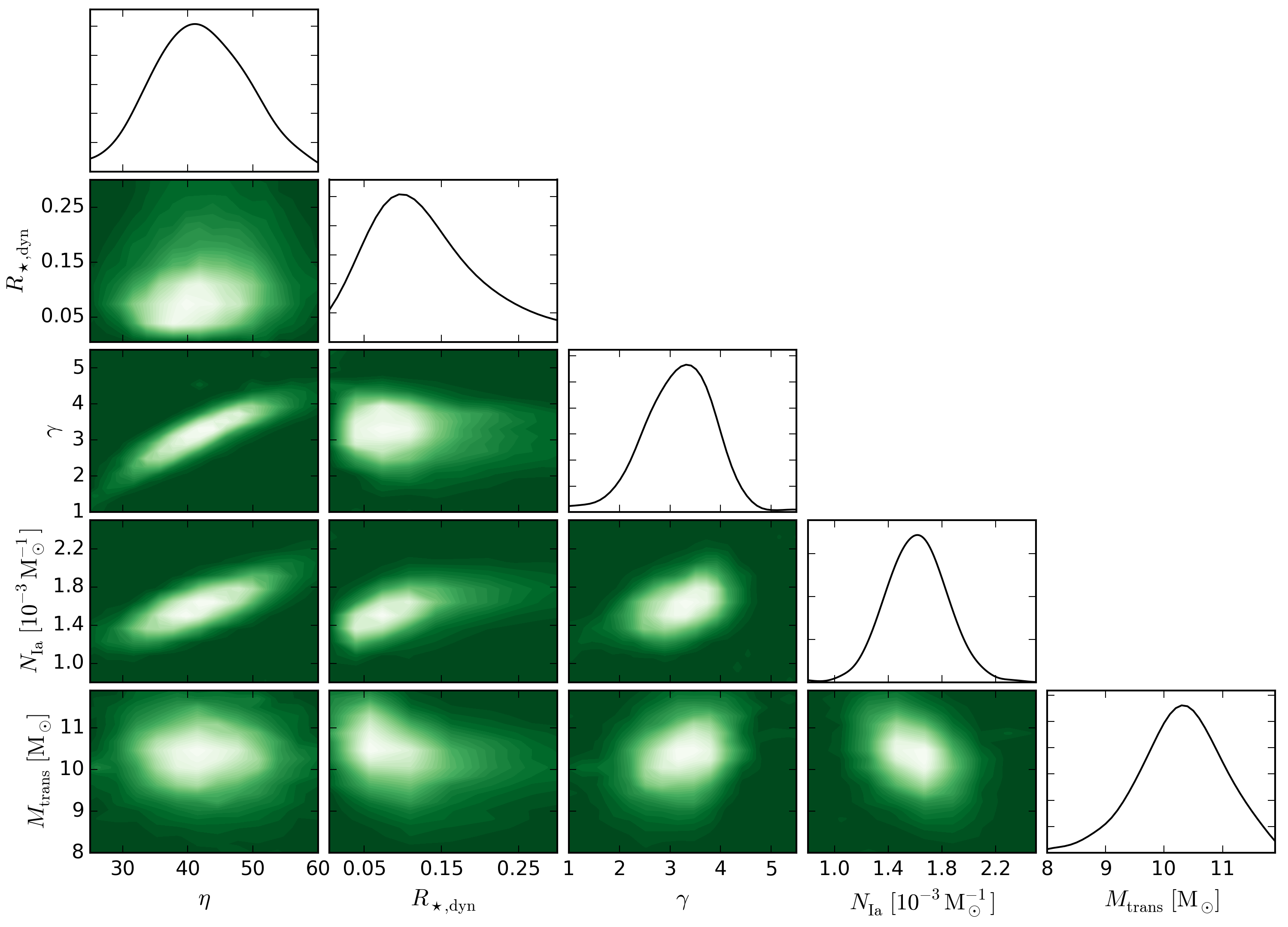}
\caption{Probability distribution of the input parameters of the MA model that provides the best fit with the chemical evolution of Sculptor.  The upper panel of each column shows the projected distribution of each parameter, where the Y axis is in arbitrary units.  The parameters are the final value of the mass-loading factor ($\eta$), the ratio between $\epsilon_\star$ and $\tau_\mathrm{dyn}$ ($R_{\star,\mathrm{dyn}}$), the mass-loading power-law index ($\gamma$), the number of SNe~Ia per stellar mass formed ($N_\mathrm{Ia}$), and the minimum mass for CC~SNe ($M_\mathrm{trans}$).}
\label{fig_MA}
\end{figure*}

At this point, even if we do not know which model is more physically valuable, we can definitely conclude that something is wrong with Cr at low [Fe/H] (see Figure~\ref{fig_fit_IO}), as the MCMC calculations failed to find a way to remove the discrepancy.  The over-production of Cr in our predictions compared to observation originates from the stellar yields used in this work, which are the same for all three models.  This demonstrates that simple chemical evolution models, such as OMEGA, are sufficient to test new sets of stellar yields and to probe nuclear astrophysics in a galactic chemical evolution context (see Section~\ref{sect_test_yields_disc}).
\\
\subsubsection{Minimum Mass for CC~SNe}
\label{sect_mmCCSNe}
The range considered for the minimum mass of CC~SNe ($M_\mathrm{trans}$), which has initially been set to cover from 8 to 12\,M$_\odot$, is probably too narrow for the MCMC calculation.  As a matter of fact, as seen in Figure~\ref{fig_IO}, the $M_\mathrm{trans}$ PDF is incomplete and should extend beyond 12\,M$_\odot$.  However, we decided not to increase this upper limit, as other works strongly indicate that 12\,M$_\odot$ is already too high for producing AGB stars (see \citealt{tww96,phwh08,smartt09,j13,f15,wh15}).  This feature occurs because the MCMC code tries to recover the best possible fit, regardless of the physical meaning of the parameters.  The unrealistic $M_\mathrm{trans}$ values resulting from the MCMC calculation indicate that important physical ingredients are missing the IO model, which is our most simplistic model.  However, this does not alter the main message of our paper (see Section~\ref{sect_ma}).

\subsection{The MA Model Results}
Figure \ref{fig_MA} shows the results of the MCMC calculation with the MA model, which is our most complex model.  The best values given by this model, for $\eta$ and $N_\mathrm{Ia}$ are different from the ones given by the IO model (see Table~\ref{tab_free_param}).  It is worth recalling, however, that the mass-loading factor is evolving with time in the MA model.  The $\eta$ parameter therefore only refers to the final value at the end of the simulation.  But, since its average value is 26.8, the IO and MA models still predict different values.  Despite these inconsistencies, these two models, as well as the SF model (see Section~\ref{sect_results_SF}), generate similar predictions when using their specific best parameters (see Figure~\ref{fig_fit_IO}). We note that the values shown for $\eta$ throughout this work are not affected by the artificial gas removal occurring during the last timesteps (see Section~\ref{sect_kssfl}).

\begin{figure}
\begin{center}
\includegraphics[width=3.3in]{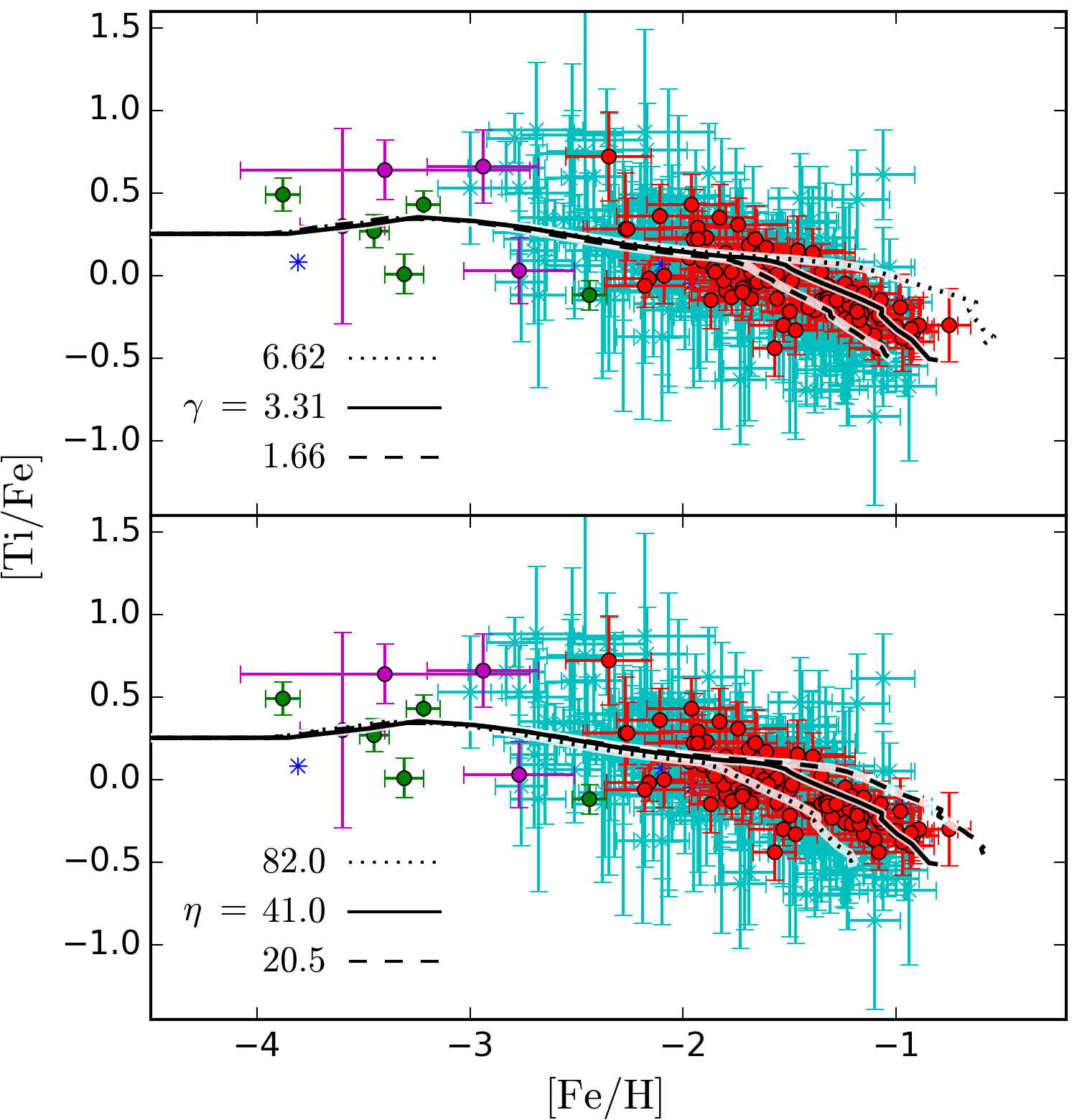}
\caption{Impact of the mass-loading power-law index (upper panel) and the strength of outflows (lower panel) on the predicted evolution of Ti as a function of [Fe/H] for the MA model.  The observational data are the same as in Figure \ref{fig_fit_IO}.}
\label{fig_gam_out_deg}
\end{center}
\end{figure}

The bump seen at [Fe/H]~$<-3$ in the predictions of alpha elements (see Section~\ref{sect_results_IO}) with the MA model (green lines) is shifted toward higher [Fe/H] values compared to the predictions of the IO model (blue lines). This is because the MA model uses a lower initial mass of gas (see $M_\mathrm{gas}$ in Table~\ref{tab_free_param}), which decreases the initial amount of hydrogen. In our models, the SFH is fixed regardless of the initial mass of gas, which always produce the same initial amount of Fe. Therefore, less gas leads to a higher [Fe/H] concentration and shifts the very-low metallicity predictions to higher [Fe/H] values. For the MCMC calculations, the location of the bump on the [Fe/H] axis has a low impact in constraining parameters, since data at [Fe/H]~$<-3$ are scarce and more uncertain compared to higher-[Fe/H] data.

As seen in Figure~\ref{fig_MA}, the $\eta$ and $\gamma$ parameters in the MA model are tightly bounded together since they both have the capacity to modify the high-metallicity end of the predicted chemical evolution (see Figure~\ref{fig_gam_out_deg}).  These two last parameters are implied in the calculation of galactic outflows and are therefore also correlated with the number of SNe~Ia, as they contribute to the iron concentration of the gas reservoir.

\subsection{The SF Model Results}
\label{sect_results_SF}
\begin{figure*}
\begin{center}
\includegraphics[width=5.6in]{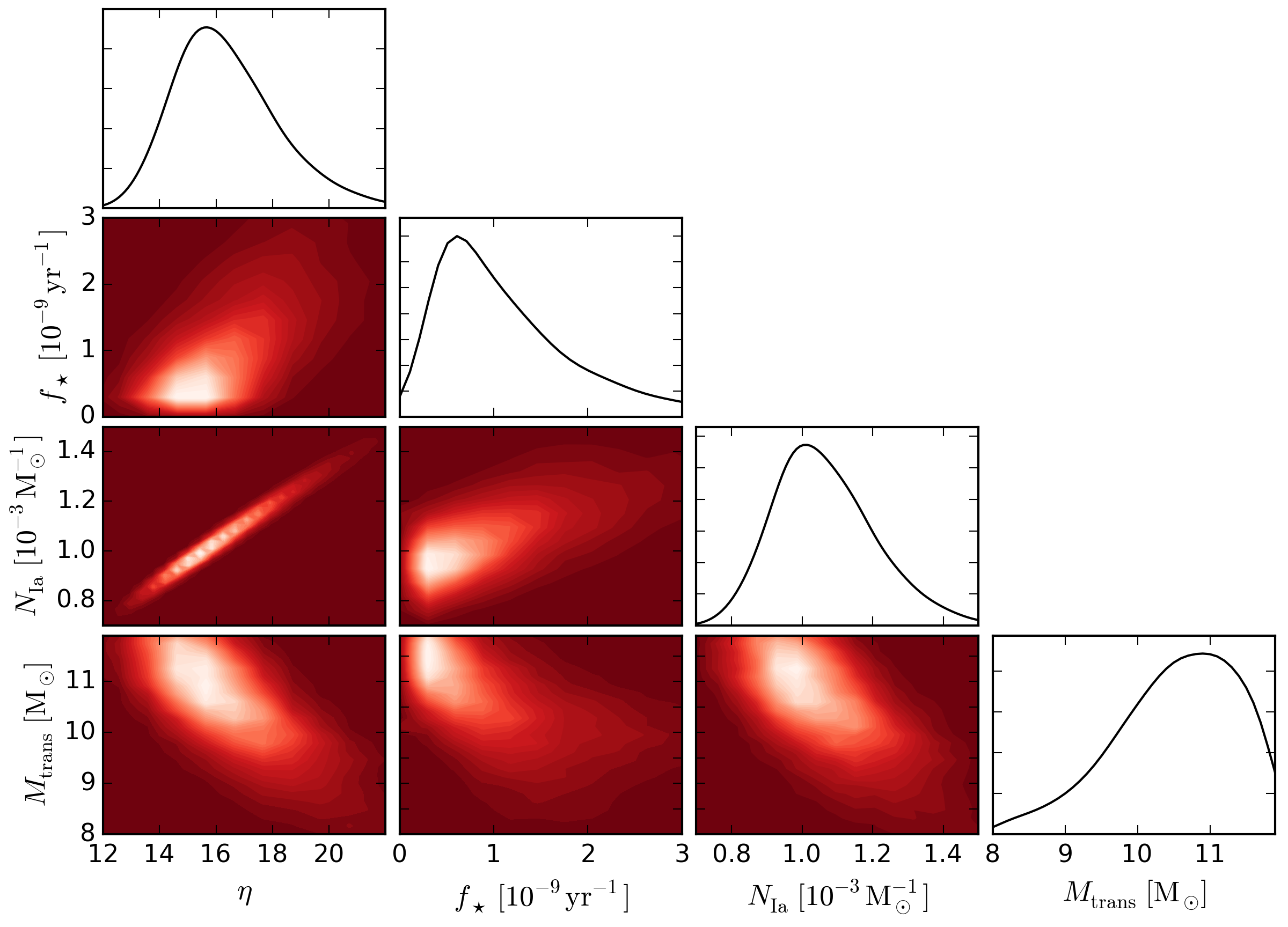}
\caption{Probability distribution of the input parameters of the SF model that provides the best fit with the chemical evolution of Sculptor.  The upper panel of each column shows the projected distribution of each parameter, where the Y axis is in arbitrary units.  The parameters are the mass-loading factor ($\eta$), the star formation efficiency ($f_\star$), the number of SNe~Ia per stellar mass formed ($N_\mathrm{Ia}$), and the minimum mass for CC~SNe ($M_\mathrm{trans}$).}
\label{fig_SF}
\end{center}
\end{figure*}

Figure~\ref{fig_SF} presents the results of the MCMC calculation with our last model, the SF model, which has only four parameters.  As seen with the MA model, there is still a degeneracy between the mass-loading factor and the number of SNe~Ia.  However, in the SF model, the correlation between these two parameters is tighter because $\eta$ and $N_\mathrm{Ia}$ are now the only parameters that significantly control the final [Fe/H] value of our simulations.

\begin{figure*}
\includegraphics[width=7in]{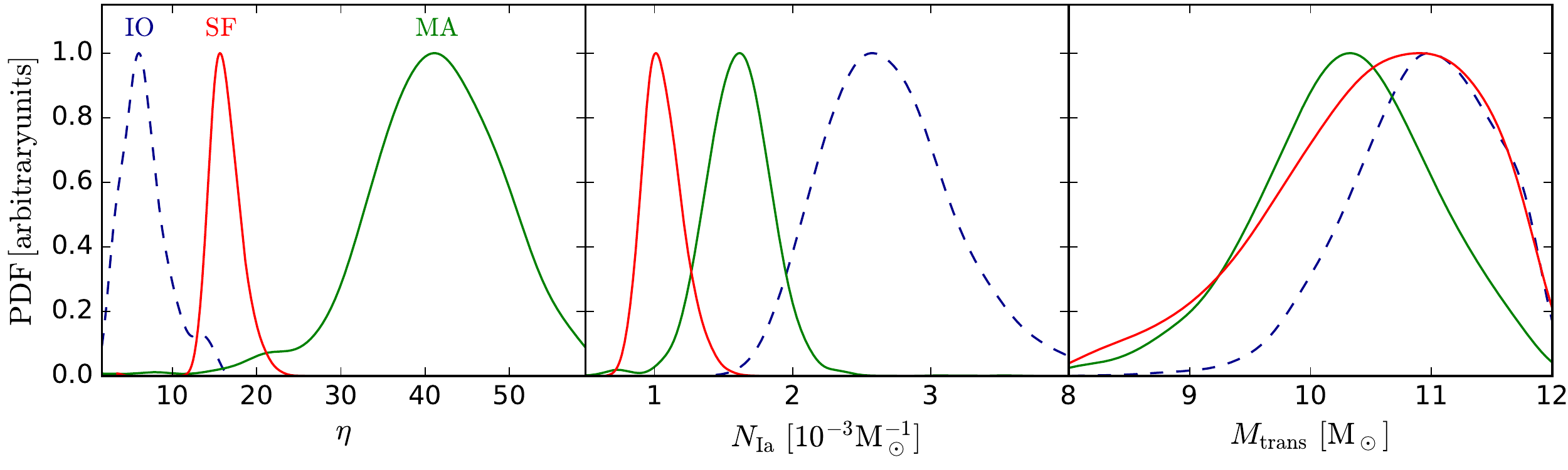}
\caption{Probability distribution functions of the input parameters that are common in the IO (blue), SF (red), and MA (green) models.}
\label{fig_model_comp}
\end{figure*}

The best fit derived with the SF model, as well as the ones derived with the IO and MA models, are shown in red in Figure~\ref{fig_fit_IO}.  The chemical evolution predictions are very similar from one model to another, and are barely distinguishable in this figure at the high-metallicity end.  However, as seen in Figure~\ref{fig_model_comp}, the best set of parameters derived by the models are different.  Indeed, besides the general agreement for the transition mass and for the average ratio between inflows and outflows for the MA and SF models (see Table~\ref{tab_free_param}), all the results are inconsistent.  But still, all three models have the same capacity to reproduce the chemical evolution of the nine selected elements in Sculptor.
\\
\section{Discussion}
In the following sections, we discuss the limitations of our study as well as the scope of simple galactic chemical evolution models.

\subsection{Free Parameters}
Some of our input parameters, such as the initial mass function, the delay-time distribution function of SNe~Ia, and the upper mass limit for CC~SNe (see \citealt{c15}), have not been included in the MCMC calculations.  This was a deliberate choice, since this work represents a first exploratory step in using MCMC with our galactic chemical evolution tools.  But, adding more parameters in the MCMC calculations would only have added more flexibility in the model to reproduce the data, and more degeneracies.  In fact, there must be a degeneracy between the mass range of CC~SNe progenitors and the initial mass function, since they both affect the amount of mass ejected by massive stars (see \citealt{f14}).  Nevertheless, the main conclusion of this paper would remain the same -- different models can fit the same data while pointing toward different sets of parameters.

\subsection{Modeling Assumptions}
\label{sect_ma}
In addition to OMEGA, there are many other galactic chemical evolution models in the literature with varying levels of complexity and modeling assumptions for inflows and outflows (see Section~\ref{sect_intro}).  We do not claim that our models are the most representative of how low-mass galaxies evolve.  In fact, because they all converge toward different parameters, at least some of our models must be \textit{wrong}.  It is also possible that all three models are insufficient to capture the complexity of galaxies.  But since the different models fit the data equally well with mutually exclusive parameters, we must conclude that a wrong model can fit the data.  We cannot distinguish between the validity of the models, and none of the output parameters, such as the mass-loading factor, can provide insight into the actual properties of Sculptor.  The observations used in this work are unable to constrain the true values of these parameters when using our three models (see Section~\ref{sect_dbm_disc}).

This raises the question of how much predictive power single-zone models such as OMEGA have in terms of galaxy evolution.  We fitted most of the observed stellar abundances, but we did not get useful information about SNe~Ia and the circulation of gas inside galaxies.  It is worth recalling that we assumed the same basic ingredients for all three models, which are galactic inflows and outflows.  The choice of ingredients included in a chemical evolution model does matter (e.g., \citealt{lm07,u15a}).  However, within the models and abundances considered here, for a given number of ingredients, there is always a way to recover equally good fits using different implementations for those ingredients.

\subsection{Number of Chemical Elements}
\label{sect_disc_num_CE}
\begin{figure}
\begin{center}
\includegraphics[width=3.3in]{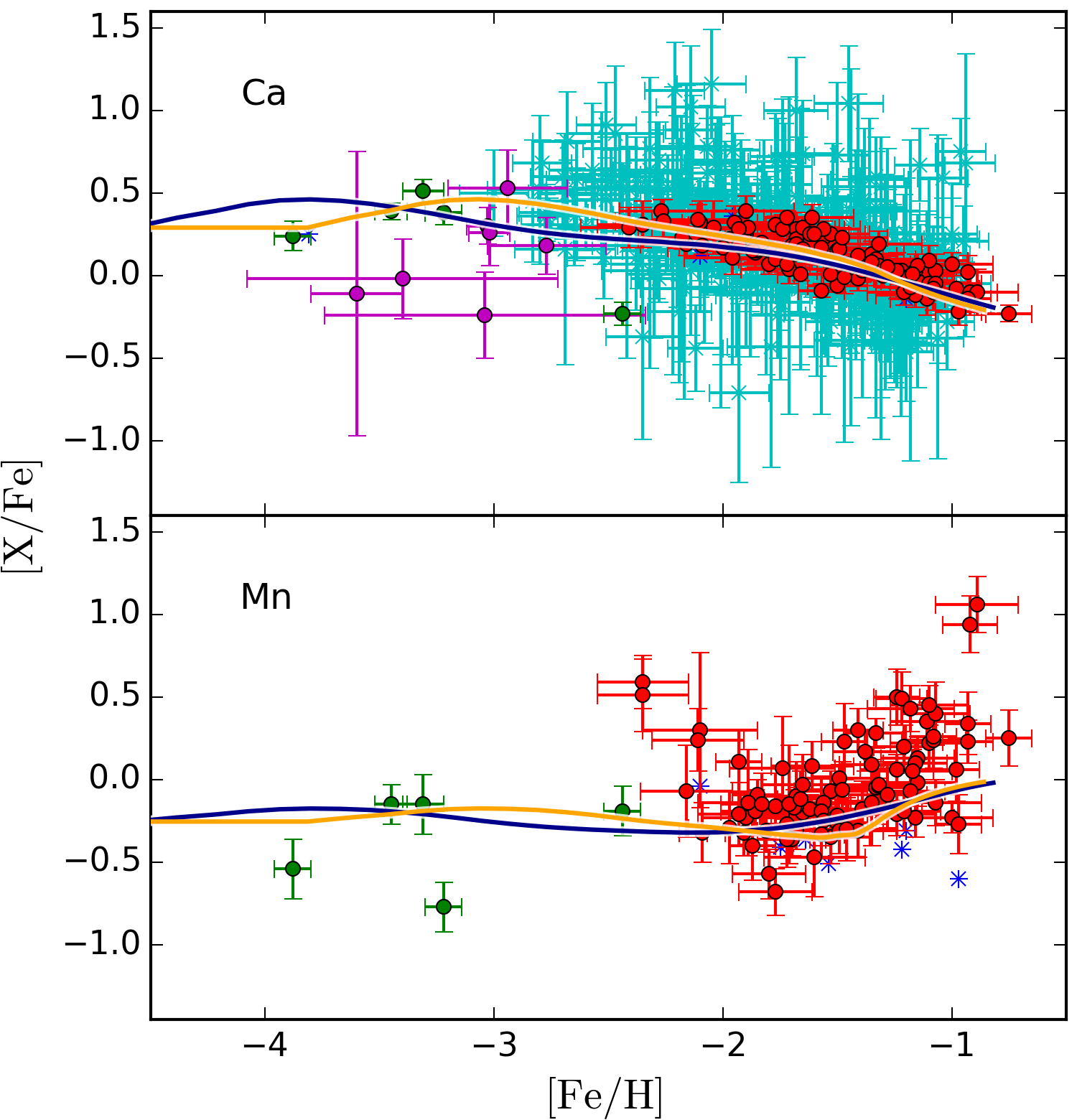}
\caption{Impact of the number of elements used in the MCMC calculation on the predicted evolution of Ca and Mn as a function of [Fe/H] for the IO model.  The blue and orange lines represent respectively the case with nine elements and with five alpha elements only.  The observational data are the same as in Figure \ref{fig_fit_IO}.}
\label{fig_nb_elem}
\end{center}
\end{figure}

This work does not represent a complete comparison between models and observations.  As noted previously, we only used nine elements to constrain our models.  The number and the choice of elements included in the MCMC calculation is important.  Adding Mn, for example, helped to constrain the parameters that control the early enrichment and the onset of SNe Ia (see also \citealt{rs13}), which are $M_\mathrm{gas}$, $R_{\star,\mathrm{dyn}}$, and $f_\star$.  As seen in Figure~\ref{fig_nb_elem}, when only the alpha elements (O, Mg, Si, Ca, and Ti) are considered in the MCMC calculation for the IO model (orange lines), the best fit for Mn is not as good as when all nine elements are included (blue lines).  We found that the best value recovered by the MCMC calculation for the $M_\mathrm{gas}$ parameter is about six times smaller when only the alpha elements are included, as opposed to when all nine elements are included.  This means that the best values presented in Table~\ref{tab_free_param} are subject to change when more elements will be added to the calculation.

Adding elements mainly ejected by low-mass stars, such as \textit{s}-process elements, would help to constrain the evolution and the strength of galactic outflows.  As a matter of fact, given the decreasing nature of the SFH of Sculptor as a function of cosmic time, galactic outflows must have been more intense at early times.  This means the elements ejected by massive stars must have been lost in greater quantity compared to \textit{s}-process elements.  It is however not clear whether adding more elements would help in determining which modeling assumptions should be discarded, as all our models seem to follow the same chemical evolution pattern (Figure~\ref{fig_fit_IO}).  More investigation is needed.
\\
\subsection{Breaking the Degeneracy Between Models}
\label{sect_dbm_disc}
A potential solution to break the degeneracy between our different models would be to include additional observational constraints that are not related to stellar abundances.  In particular, we could use the observed current star formation efficiency, as all of our models have different predictions for that quantity (see Table~\ref{tab_free_param}).  But unfortunately, dwarf spheroidal galaxies such as Sculptor no longer form stars, which prevents us from using that constraint.  In the nearby galaxy sample of \cite{l08}, dwarf irregular galaxies have a current specific star formation efficiency of roughly between 10$^{-11}$ and 10$^{-9}$~yr$^{-1}$.  Although our models are broadly consistent with those observations (see also \citealt{lm04,vmvl14}), we cannot use those values as direct constraints, since they represent the current state of gas-rich systems as opposed to the past history of one specific spheroidal galaxy.

By considering the current gas fraction of dwarf spheroidal galaxies, which is essentially zero, we could already think about discarding the IO model, since it is the only model that does not lose its entire gas reservoir.  But on the other hand, we could add to the model a prescription that artificially empties the gas reservoir near the end of the simulations to mimic tidal or ram pressure stripping.  After all, the removal of gas in the MA and SF models is also rather artificial.  Given the current degeneracy between the models in fitting the stellar abundances, and the fact that our three models predict different behaviours for the gas, we believe that gas-rich systems like the Milky Way or dwarf irregular galaxies would be better targets to model in order to discard some of our modeling assumptions.  Those galaxies would provide specific observational constrains for the galactic outflow rate, the mass fraction of gas, and the star formation efficiency, for which all three models predict different quantities.  But unfortunately, it is currently challenging to derive observationally the chemical abundances of old stars in dwarf irregular galaxies (\citealt{v03,t09}).

In this paper, we only considered one galaxy as an exploratory step toward a broader use of MCMC calculations in chemical evolution studies. Applying the MCMC approach to several local galaxies simultaneously would help to better constrain our parameters and determine which galaxy models should be discarded. This will be addressed in a forthcoming paper.

\subsection{Additional Physical Ingredients}
\label{sect_disc_add_phy_ing}
Several physical processes, such as metal-rich outflows, pre-enriched inflows, star formation thresholds, and gas stripping, have been neglected in our models.  This could affect the best values of our input parameters derived by the MCMC calculations.

If metal-rich outflows were included, a larger quantity of metals ejected by CC~SNe would be lost from the system (e.g., \citealt{y13,cmd15}), which could reduce the predicted [$\alpha$/Fe] ratios. In that case, the MCMC calculations could reduce the strength of galactic outflows or the number of SNe~Ia in order to either retain more alpha elements or reduce the iron production.  If pre-enriched inflows were considered, there would be more metals in the system and our predictions could be shifted toward higher [Fe/H] values (e.g., \citealt{andrews16}).  To counterbalance this effect and to recover our fits, the strength of galactic outflows or the mass of the gas reservoir should be increased in order to either eject more metals or dilute the [Fe/H] concentration. However, pre-enriched inflows are more important for more massive galaxies (\citealt{brook14}).

If a threshold mass of gas below which no star formation can occur was introduced in the SF or MA models, more gas would be required to form the same amount of stars (e.g., \citealt{croton06,sd14}). To recover the lower gas content originally needed to fit the data with our models, the star formation efficiency would have to be increased. We did not directly consider gas stripping processes in our equations.  However, the gas reservoir in the MA and SF models is emptied at the end of our simulations, which indirectly mimic the impact of gas stripping in the context of one-zone models. During the active star-forming period, if gas stripping was included in addition to galactic outflows, more metals would be removed from the systems and replaced by primordial gas.  To recover the balance needed in the gas circulation to reproduce observations, the strength of galactic outflows would have to be reduced.

The best set of input parameters shown in Table~\ref{tab_free_param} should therefore be considered with caution (see also Section~\ref{sect_disc_num_CE}).  Our study should be considered as an experiment that highlights the capacity of different models to fit a certain collection of stellar abundances.  Within this context, having more ingredients and therefore more parameters would simply add more flexibility to fit the stellar abundances, which may not help to break the degeneracy between our three models, as opposed to adding more observational constraints (see Section~\ref{sect_dbm_disc}). Including more ingredients is generally necessary to solve discrepancies between models and observations (e.g., \citealt{henri13}), which typically emerge when more constraints are considered.  But in our case, we do not expect that new galaxy evolution ingredients would solve the Cr discrepancy (see Section~\ref{sect_res_deg_bet_mod}), as this feature is a direct consequence of our choice of input stellar yields.

That conclusion, however, does not mean that more complex chemical evolution simulations are not needed.  Simple models such as OMEGA are designed to reproduce global trends and cannot simulate other interesting aspects such as abundance gradients (e.g., \citealt{cmr01,cmfc07,mcm14a}), non-uniform mixing and dispersion (e.g., \citealt{m07,m08,p12,k14,hirai15,r15,wpt15,r16,yqj15}), and the chemical signatures of galaxy mergers (e.g., \citealt{k04,r10,rkb10,mcm14b,rl15}).

\subsection{Tests for Stellar Models}
\label{sect_test_yields_disc}
Stellar yields are the foundation of all chemical evolution simulations, from one-zone models to cosmological hydrodynamical simulations of galaxy evolution.  Once a new set of yields is calculated, it is necessary to proceed to a consistency check before publishing the results or before introducing them into time-consuming simulations.  Because of its low computational cost, a one-zone model such as OMEGA is ideal to accomplish this task.  In this paper, we have shown that having a degeneracy between our models is problematic for understanding how galaxies evolve.  However, this degeneracy turns out to be a convenient result for testing stellar yields.

Observational data can be fitted equally well with multiple chemical evolution models.  We found that using a \textit{complex} implementation (e.g, the MA model) gives similar results to the most simplistic implementation (e.g., the IO model).  This suggests that simple models such as OMEGA can be used to perform valuable consistency checks of stellar yields.  For this purpose, we do not need to use a complex chemical evolution model, and therefore we do not need to worry about the impact of the choice of model.

\section{Conclusions}
The first goal of this paper was to present our simple galactic chemical evolution code, OMEGA, which is part of our numerical pipeline designed to connect nuclear physics and stellar evolution with galactic chemical evolution.  The second goal was to evaluate the impact of using different modeling assumptions in simple chemical evolution studies.  To do so, we introduced the three different implementations of OMEGA (the IO, MA, and SF models) into an MCMC calculation in order to reproduce the abundance evolution of O, Mg, Si, Ca, Ti, Cr, Mn, Ni, and Co observed in Sculptor, a dwarf spheroidal galaxy.  Using this approach, we could find the best set of parameters along with their confidence levels, and highlight the degeneracy between the different input parameters, for each model.

We found that all three models are able to reproduce to the same degree the stellar abundance trends observed in Sculptor (see also \citealt{rs13}), although they use different modeling assumptions.  However, the values found by the MCMC calculations for the strength of galactic outflows and the number of SNe~Ia per stellar population, are generally different from one model to another.  With a limited number of observations, having a good fit does not necessarily mean that the input parameters are relevant.

Our experiment suggests that studying chemical evolution with nine elements and simple models is not sufficient to learn about galaxy evolution.  More elements or additional constraints that are not associated with stellar abundances are needed in order to distinguish between the different modeling assumptions.  In that regard, dwarf spheroidal galaxies are probably not the best targets, as they do not provide any constraint for the star formation efficiency and the global circulation of gas.  However, from a stellar evolution perspective, our results are encouraging since they show that chemical evolution predictions are predominantly driven by stellar yields.  Simple models such as OMEGA are therefore sufficient to test and validate new stellar models, at least as a first order approximation.

\section*{acknowledgments}
We are thankful to Vanessa Hill for valuable discussions on the chemical evolution of Sculptor and access to the data in advance of publication.  We are thankful to Brad Gibson for instructive and useful conversations regarding chemical evolution simulations.  The MCMC calculations have been performed on the Mammouth parallele 2 supercomputer associated with Calcul Quebec and Calcul Canada.  This research is supported by the National Science Foundation (USA) under Grant No. PHY-1430152 (JINA Center for the Evolution of the Elements), and by the FRQNT (Quebec, Canada) postdoctoral fellowship program.  BWO was supported by the National Aeronautics and Space Administration (USA) through grant NNX12AC98G and Hubble Theory Grant HST-AR-13261.01-A. He was also supported in part by the sabbatical visitor program at the Michigan Institute for Research in Astrophysics (MIRA) at the University of Michigan in Ann Arbor, and gratefully acknowledges their hospitality. FH acknowledges support through a NSERC Discovery Grant (Canada).

\label{lastpage}

\end{document}